# Infinite Body Centered Cubic Network of Identical Resistors

**J. H. Asad[1,\*], A. A. Diab[2], R. S. Hijjawi[3], and J. M. Khalifeh[4]**
[1]Dep. of Physics- Tabuk University- P. O Box 741- Tabuk 71491- Saudi Arabia
[2]General Studies Department- Yanbu Industrial College- P. O Box 30436- Yanbu Industrial City- KSA
[3]Dep. of Physics- Mutah University- Karak- Jordan
[4]Dep. of Physics-Jordan University- Amman 11942 - Jordan

## Abstract

We express the equivalent resistance between the origin $(0,0,0)$ and any other lattice site $(n_1, n_2, n_3)$ in an infinite Body Centered Cubic (BCC) network consisting of identical resistors each of resistance $R$ rationally in terms of known values $b_o$ and $\pi$. The equivalent resistance is then calculated. Finally, for large separation between the origin $(0,0,0)$ and the lattice site $(n_1, n_2, n_3)$ two asymptotic formulas for the resistance are presented and some numerical results with analysis are given.

**\* Corresponding author e-mail: jhasadd1@yahoo.com.**

# 1. Introduction

The Lattice Green's Function (LGF) is a basic concept in physics. Many quantities of interest in solid-state physics can be expressed in terms of it. For example, statistical model of ferromagnetism such as Ising model [1], Heisenberg model [2], spherical model [3], lattice dynamics [4], random walk theory [5, 6], and band structure [7, 8]. In Economou's book [9] one can find an excellent introduction to the LGF, where a review of the LGF of the so-called tight-binding Hamiltonian (TBH) used for describing the electronic band structures of crystal lattices is presented. The LGF defined in this paper is related to the GF of the TBH. Many efforts have been paid on studying the LGF of cubic lattices [10-25].

The LGF for the BCC lattice has been expressed as a sum of simple integrals of the complete elliptic integral of the first kind [10], Morita and Horiguci [11] presented formulas which are convenient for the evaluation of the LGF for the Face Centered Cubic (FCC), BCC and rectangular lattices. These formulas involve the complete elliptic integral of the first kind with complex modulus. Morita [12] derived a recurrence relation, which gives the values of the LGF along the diagonal direction from a couple of the elliptic integrals of the first and second kind for the square lattice with discussions of how to apply the result to the BCC lattice. Finally, Glasser and Boersma [23] expressed the values of the LGF of the BCC lattice rationally. One can find more works in these works and references within them.

The calculation of the equivalent resistance in infinite networks is a basic problem in the electric circuit theory. It is of so interest for physicists and electrical engineering. There are mainly three approaches used to solve such a problem. The superposition of current distribution has been used to calculate the effective resistance between adjacent sites on infinite networks [26-28].

A mapping between random walk problems and resistor networks problems have been used by Monwhea Jeng [29], where his method was used to calculate the effective resistance between any two sites in an infinite two-dimensional square lattice of unit resistors.

A third educational important method based on the Lattice Green's Function (LGF) of the lattices has been used in calculating the equivalent resistance [30-38]. This method has been applied to both perfect and perturbed square, simple cubic (SC) networks and recently to the FCC network

The present work is organized as follows:

In Sec. 2 we briefly introduced the basic formulas of interest for the LGF of the BCC network. In Sec. 3 an application to the LGF of the BCC network has been applied to express the equivalent resistance between the origin and the lattice site $(n_1,n_2,n_3)$ in the infinite BCC network rationally in terms of some constants, and the asymptotic behavior for the resistance is also investigated as the separation between the two sites goes to infinity. Finally, we close this paper (Sec. 4) with a discussion for the results obtained.

## 2. Basic Definitions and Preliminaries

The LGF for the BCC lattice appears in many areas in physics (e.g., Ising model [1, 39-40], Heisenberg model [41- 43] and spherical models [44-46] and it is defined as [13, 23] :

$$B(E;n_1,n_2,n_3) = \frac{1}{\pi^3}\int_0^\pi\int_0^\pi\int_0^\pi \frac{Cos(n_1 u)Cos(n_2 v)Cos(n_3 w)}{E - CosuCosvCosw}dudvdw. \quad (1)$$

Where $E \geq 1$

$n_1, n_2,$ and $n_3$ are either all even or all odd integers.

The LGF for the BCC lattice at the site $(0,0,0)$ which represents the origin of the lattice for $E = 1$ (i.e., $B(1;0,0,0) = b_o$) was of so interests in physics and it was carried out first by Van Peijpe [47] and later on by Watson [48]. They showed that:

$$b_o = B(1;0,0,0) = \frac{4}{\pi^2}[K(\frac{1}{\sqrt{2}})]^2 = \frac{\Gamma^4(\frac{1}{4})}{4\pi^3} = 1.3932039297. \quad (2)$$

Where $K$ is the complete elliptic integral of the first kind, and $\Gamma$ is the Gamma function

In a recent work the LGF for the infinite BCC lattice has been expressed rationally as [23]:

$$B(1;n_1,n_2,n_3) = \sigma_1 b_o + \frac{\sigma_2}{\pi^2 b_o} + \sigma_3. \quad (3)$$

Where $\sigma_1$, $\sigma_2$ and $\sigma_3$ are rational numbers.

# 3. Application: Evaluation of the resistance $R(n_1,n_2,n_3)$ in an Infinite BCC Network

The aim of this section is to express the equivalent resistance between the origin $(0,0,0)$ and the lattice site $(n_1,n_2,n_3)$ in the infinite BCC network which is consisting from identical resistors rationally in terms of $b_o$ and $\pi$.

First of all, it has been showed that for a 3D infinite network consisting of identical resistors each of resistance $R$, the equivalent resistance between the origin and any other lattice site is [30]:

$$R(r) = 2[G(0) - G(r)]. \qquad (4)$$

Where $r$ is the position vector of the lattices point, and for a d-dimensional lattice it has the following form:

$$r = n_1 a_1 + n_2 a_2 + \ldots + n_d a_d. \qquad (5)$$

With $n_1, n_2, \ldots, n_d$ are integers, and $a_1, a_2, \ldots, a_d$ are independent primitive translation vectors.

Also, the equivalent resistance between the origin and any other lattice site is can be expressed in an integral form as [30]:

$$R(n_1, n_2, \ldots, n_d) = R \int_{-\pi}^{\pi} \frac{dx_1}{2\pi} \ldots \int_{-\pi}^{\pi} \frac{dx_d}{2\pi} \frac{1 - \exp(in_1 x_1 + in_2 x_2 + \ldots + in_d x_d)}{\sum_{i=1}^{d}(1 - \cos x_i)}.$$

(6)

On the other hand, the LGF for a 3D hypercube read as [30]:

$$G(n_1, n_2, \ldots, n_d) = \int_{-\pi}^{\pi} \frac{dx_1}{2\pi} \ldots \int_{-\pi}^{\pi} \frac{dx_d}{2\pi} \frac{\exp(in_1 x_1 + in_2 x_2 + \ldots + in_d x_d)}{2\sum_{i=1}^{d}(1 - \cos x_i)}.$$

(7)

For cubic lattices $d = 3$. Then substituting $d = 3$ into Eqs. (6) and (7) and comparing them with Eq. (4) one get:

$$R(n_1, n_2, n_3) = R[b_o - B(1; n_1, n_2, n_3)].$$

(8)

Now make use of Eq. (3) and Eq. (8) one yields

$$\frac{R(n_1,n_2,n_3)}{R} = r_1 b_o + \frac{r_2}{\pi^2 b_o} + r_3. \tag{9}$$

Where $r_1 = 1-\sigma_1$, $r_2 = -\sigma_2$ and $r_3 = -\sigma_3$ are rational numbers. These rational numbers, for the sites from $(0,0,0)$ to $(8,8,8)$, can be gathered from **[13, appendix A]**. In Table 1 below we present these rational numbers.

Based on the recurrence formula presented in **[13, Eq. (5.8)]** we have calculated additional rational values for the sites from $(9,1,1)$ to $(10,0,0)$ and arranged them in Table 1 below.

Since the LGF is an even function (i.e., $B(1;n_1,n_2,n_3) = B(1;-n_1,-n_2,-n_3)$) and due to the fact that the infinite BCC network is pure and symmetric, then as a result $R(n_1,n_2,n_3) = R(-n_1,-n_2,-n_3)$.

Finally, it is interesting to study the asymptotic behavior of the equivalent resistance for large separation between the origin $(0,0,0)$ and any other lattice site $(n_1,n_2,n_3)$.

The asymptotic form of $B(1;0,0,n_3)$, as $n_3 \to \infty$, is given as **[13]**:

$$B(1;0,0,2n_3) \to \frac{1}{\pi n_3}(1 - \frac{1}{8n_3^2} + \frac{1}{128n_3^4} - \frac{173}{1024n_3^6}). \tag{10}$$

While, for large value of $|n| = \sqrt{n_1^2 + n_2^2 + n_3^2}$ it has been shown that **[13]** $B(1;n_1,n_2,n_3)$ has the following asymptotic formula

$$B(1;n_1,n_2,n_3) \approx \frac{2}{\pi}|n|^{-1}[1 - \frac{9}{8}|n|^{-2} + \frac{5}{8}|n|^{-6}(n_1^4 + n_2^4 + n_3^4) +$$

$$\frac{15}{4}|n|^{-6}(n_1^2 n_2^2 + n_2^2 n_3^2 + n_3^2 n_1^2)]. \tag{11}$$

Inserting Eq. (10) and Eq. (11) into Eq. (8), one gets the following two equations:

$$\frac{R_o(0,0,2n_3)}{R} \to b_o - \frac{1}{\pi n_3}(1 - \frac{1}{8n_3^2} + \frac{1}{128n_3^4} - \frac{173}{1024n_3^6}). \tag{12}$$

$$\frac{R_o(n_1,n_2,n_3)}{R} \to b_o - \frac{2}{\pi}|n|^{-1}[1 - \frac{9}{8}|n|^{-2} + \frac{5}{8}|n|^{-6}(n_1^4 + n_2^4 + n_3^4) -$$

$$\frac{15}{4}|n|^{-6}(n_1^2 n_2^2 + n_2^2 n_3^2 + n_1^2 n_3^2)]. \tag{13}$$

The last asymptotic formula agrees with Eq. (12) for $n_1 = 0$, $n_2 = 0$ and for $n_3 = 2n_3$. In addition, the above two asymptotic formulas can be used to check the results obtained in Table 1 below. For example,

$$\frac{R_o(8,0,0)}{R} \cong 1.31425;$$

$$\frac{R_o(8,8,6)}{R} \cong 1.34413;$$

$$\frac{R_o(8,8,8)}{R} \cong 1.34778;$$

$$\frac{R_o(9,9,9)}{R} \cong 1.35273;$$

$$\frac{R_o(10,0,0)}{R} \cong 1.32986. \tag{15}$$

From the above two asymptotic formulas, one can see that as $n_3 \to \infty$, or as $|n| \to \infty$ then the resistance goes to a finite value (i.e., goes to $b_o$).

## 4. Results and Discussion

In this work we have expressed the equivalent resistance between the origin $(0,0,0)$ and any other lattice site $(n_1, n_2, n_3)$ in an infinite BCC network consisting of identical resistors each of resistance $R$ rationally in terms of the two known values $b_o$ and $\pi$. The rational number $r_1$, $r_2$ and $r_3$ presented in Eq. (14) were calculated using some recurrence formulas. In Fig. 1 and Fig. 2 the equivalent resistance is plotted against the lattice site.

Figure 1 shows the resistance in an infinite BCC lattice against the site $(n_1, n_2, n_3)$ along the [100] direction. From this figure it is clear that the resistance is symmetric.

Figure 2 shows the resistance in an infinite BCC lattice against the site $(n_1, n_2, n_3)$ along the [111] direction. From this figure it is clear that the resistance is symmetric.

The above figures indicate that as the separation between the origin and the lattice site $(n_1, n_2, n_3)$ increases, then the equivalent resistance approaches a finite value (i.e., $b_o = 1.3932039297$) as explained above.

Finally, it is worth mention that for the case of Face Centered Cubic (FCC) network the equivalent resistance as the separation between the origin and any other lattice site approaches a finite value ($f_o = 0.4482203944$) **[37]** where $f_o$ is the LGF at the origin for the FCC lattice, while for the case of an infinite SC network it goes to the finite value $g_o = 0.505462$ **[30, 33]**, where $g_o$ is the LGF at the origin for the SC lattice. Whereas for the infinite square lattice it goes to infinity **[30, 34]**

# Table Captions

**Table 1:** Values for selected rational numbers $r_1$, $r_2$, $r_3$ and $R(n_1,n_2,n_3)$ for sites $(0,0,0)$ to $(10,0,0)$

**Table 1:**

| $(n_1,n_2,n_3)$ | $r_1$ | $r_2$ | $r_3$ | $R(n_1,n_2,n_3)/R$ |
|---|---|---|---|---|
| 000 | 0 | 0 | 0 | 0 |
| 111 | 0 | 0 | 1 | 1.0000 |
| 002 | 1 | -4 | 0 | 1.1023 |
| 022 | 0 | 16 | 0 | 1.16360 |
| 222 | -3 | -36 | 8 | 1.20228 |
| 113 | 2 | -8 | -1 | 1.20461 |
| 133 | -4 | 80 | 1 | 1.24521 |
| 333 | -18 | -504 | 63 | 1.26877 |
| 004 | 8/9 | 0 | 0 | 1.23840 |
| 024 | 25/9 | -36 | 0 | 1.25190 |
| 224 | 104/9 | 16 | -16 | 1.26285 |
| 044 | -112/9 | 256 | 0 | 1.28003 |
| 244 | -407/9 | 444 | 32 | 1.28626 |
| 444 | -360/9 | -5376 | 448 | 1.30059 |
| 115 | -2/9 | 8 | 1 | 1.27220 |
| 135 | 120/9 | -224 | -1 | 1.28558 |
| 335 | 810/9 | 920 | -191 | 1.29564 |
| 155 | -652/9 | 1392 | 1 | 1.30374 |
| 355 | -4266/9 | 1192 | 575 | 1.30990 |
| 555 | 7650/9 | -48840 | 2369 | 1.31934 |
| 006 | 1 | -36/25 | 0 | 1.28848 |
| 026 | -16/9 | 1296/25 | 0 | 1.29327 |
| 226 | -155/9 | 444/25 | 24 | 1.29753 |
| 046 | 409/9 | -21316/25 | 0 | 1.30487 |
| 246 | 1112/9 | -42224/25 | -48 | 1.30795 |
| 446 | 5481/9 | 379804/25 | -1952 | 1.31569 |
| 066 | -288 | 138384/25 | 0 | 1.31802 |
| 266 | -5147/9 | 249596/25 | 72 | 1.31998 |
| 466 | -165600/36 | -563056/25 | 8048 | 1.32510 |
| 666 | 169317/9 | -9083844/25 | 216 | 1.33175 |
| 117 | 20/9 | -272/25 | -1 | 1.30476 |
| 137 | -808/36 | 10856/25 | 1 | 1.31055 |
| 337 | -7616/36 | -29888/25 | 383 | 1.31541 |
| 157 | 9480/36 | -125320/25 | -1 | 1.31962 |
| 357 | 12620/9 | -275440/25 | -1151 | 1.32317 |
| 557 | 20584/9 | 4757600/25 | -17025 | 1.32903 |
| 177 | -57824/36 | 769376/25 | 1 | 1.32912 |
| 377 | -207128/36 | 1964312/25 | 2303 | 1.33153 |
| 577 | -1396840/36 | -2848376/5 | 95489 | 1.33566 |
| 777 | 9331056/36 | -42996912/25 | -236033 | 1.34058 |
| 008 | 1664/1764 | 0 | 0 | 1.31422 |
| 028 | 8164/1764 | -1764/25 | 0 | 1.31641 |
| 228 | 48256/1764 | -1648/25 | -32 | 1.31845 |
| 048 | -183136/1764 | 50176/25 | 0 | 1.32213 |
| 448 | -3604288/1764 | -737024/25 | 4992 | 1.32815 |
| 068 | 2029540/1764 | -550564/25 | 0 | 1.32949 |
| 268 | 3543104/1764 | -928496/25 | -96 | 1.33070 |
| 468 | 28134948/1764 | -664004/25 | -20288 | 1.33398 |
| 668 | -38421504/1764 | 49955984/25 | -114976 | 1.33849 |
| 088 | -12686080/1764 | 3444736/25 | 0 | 1.33687 |
| 288 | -19609372/1764 | 5280412/25 | 128 | 1.33772 |
| 488 | -115817696/1764 | 13932544/25 | 50944 | 1.34006 |
| 688 | -436242204/1764 | -216804644/25 | 975232 | 1.34340 |
| 888 | 5048578944/1764 | 165654528/25 | -4469248 | 1.34721 |
| 119 | -148/441 | 272/25 | 1 | 1.32369 |
| 139 | 17230/441 | -17912/25 | -1 | 1.32666 |
| 339 | 19368/49 | 30816/25 | -639 | 1.3293 |
| 159 | -306598/441 | 66616/5 | 1 | 1.33168 |
| 359 | -1425500/441 | 177776/5 | 1919 | 1.33381 |
| 559 | -6683968/441 | -475264 | 55681 | 1.33751 |
| 179 | 41833/63 | -3179392/25 | -1 | 1.33754 |
| 379 | 1200158/63 | -7803464/25 | -3839 | 1.33914 |
| 579 | 10571410/63 | 4255912/5 | -295681 | 1.34199 |
| 779 | -678572 | 435693424/25 | -322047 | 1.34551 |
| 199 | -17834440/441 | 19368352/25 | 1 | 1.3433 |
| 399 | -4533062/49 | 42106968/25 | 6399 | 1.34446 |
| 599 | -335991178/441 | 10935752/5 | 902401 | 1.34657 |
| 799 | -2907724/7 | -2645823088/25 | 8275455 | 1.34938 |
| 999 | 1270018116/49 | 7998622128/25 | -59378175 | 1.35244 |
| 0010 | 1 | -196/225 | 0 | 1.32985 |

# Figure Captions

**Fig.1:** Resistance between the origin $(0,0,0)$ and the site $(n,0,0)$ along [100] direction for BCC network.

**Fig.2:** Resistance between the origin $(0,0,0)$ and the site $(n,n,n)$ along [111] direction for BCC network.

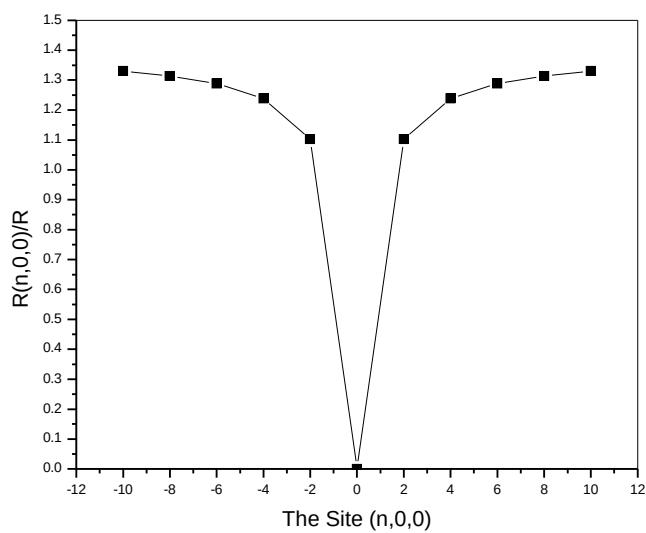

**Fig.1**

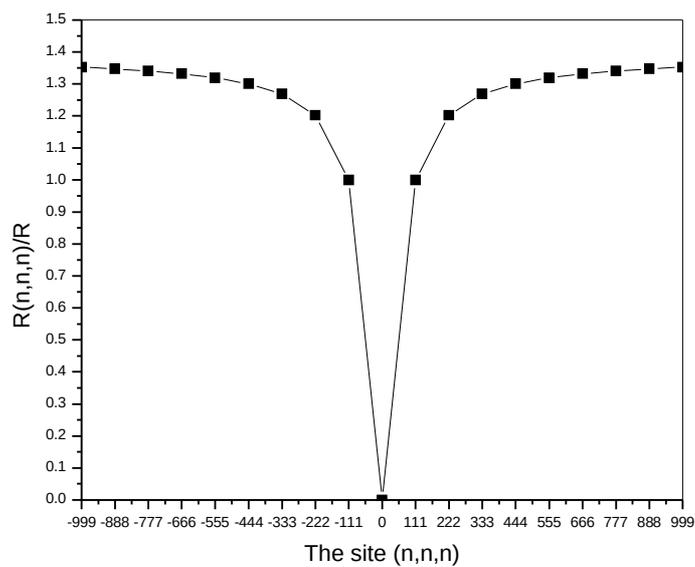

**Fig.2**

**Reference:**

[1] Brout, R. (1960). Physical Review 118, 1009.

[2] Dalton, N. W. and Wood, D. W. (1967). Critical behavior of the simple anisotropic Heisenberg model. Proceedings of the Physical Society (London) 90, 4591.

[3] Tax, M. (1955). Physical Review 97, 629.

[4] Dederichs, P. H., Schroeder, K., and Zeller, R. (1980). Point Defect in Metals II, Springer, Berlin.

[5] Hughes, B. D. (1986). On returns to the starting site in Lattice random walk. Physica A 134, 443.

[6] Montroll, E. W. (1956). Random walks in multidimensional spaces especially on periodic lattice. Journal of the Society of Industrial Applied Mathematics 4, 241.

[7] Koster, G. F. and Slater, D. C. (1954). Simplified impurity calculation. Physical Review 96, 1208.

[8] Li, Q., Soukoulis, C., Economou, E. N., and Grest, G. S. (1989). An isotropic tight-binding model for localization. Physical Review B 40, 2825.

[9] E. N. Economou, Green's functions in Quantum Physics (Springer-Verlag, Berlin, Germany, 1983), 2nd ed., pp. 3–18, pp. 71–96.

[10] Tohru Morita and Tsuyoshi Horiguchi (1971). Lattice Green's Functions for the Cubic Lattices in Terms of the Complete Elliptic Integral. Journal of Mathematical Physics 12, 981

[11] Morita, T. and Horiguci, T. (1971). Calculation on the lattice Green's function for the BCC, FCC, and rectangular lattices. Journal of Mathematical Physics 12, 986.

[12] Morita, T. (1971). Useful procedure for computing the lattice Green's function, square, tetragonal, and BCC lattices. Journal of Mathematical Physics 12, 1744.

[13] Joyce, G. S. (1971). Exact results for body centered cubic lattice Green's function with applications in lattice statistics I. Journal of Mathematical Physics 23, 1390.

[14] Joyce, G. S. (1971). Exact evaluation of the body centered cubic lattice Green function. Journal of Physics C: Solid State Physics 23, 1510.

[15Horiguchi, T. (1971). *Journal of Physics Society Japan* **30**, 1261.

[16] Katsura, S. and Horiguchi, T. (1971). Lattice Green's function for the body centered cubic Lattice. Journal of Mathematical Physics 12, 230.

[17] Glasser, M. L. (1972). *Journal of Mathematical Physics* **13**(8), 1145.

[18] Inoue, M. (1975). Lattice Green's function for the body centered cubic lattice. Journal of Mathematical Physics 16, 809.

[19] Mano, K. (1975). Remarks on the Green's function for cubic lattices. Journal of Mathematical Physics 16, 1726.

[20] Morita, T. (1975). *Journal of Physics A: Mathematics and General* **8**, 478.

[21] Morita, T. and Horiguchi, T. (1975). *Journal of Physics C: Solid State Physics* **8**, L232.

[22] Glasser, M. L. and Zuker, I. J. (1977). *Proceedings of National Academics of Science USA*, **74**, 1800.

[23] Glasser, M. L. and Boersma, J. (2000). *Journal of Physics A: Mathematics and General* **33(**28), 5017.

[24] Hijjawi R. S, Asad J. H, Sakaji A. J, and Khalifeh J. M. (2004). Lattice Green's Function for the Face Centered Cubic Lattice. Int. J. Theo. Phys., (43)11: 2299-2309.

[25] Asad J. H. (2007). Differential Equation Approach for One- and Two-Dimensional Green's Function. Mod. Phys. Letters B, (21) 2-3. 139.

[26] F. J. Bartis, Am. J. Phys. **35** (1967) 354–362.

[27] G. Venezian, Am. J. Phys. **62** (1994) 1000–1004.

[28] D. Atkinson and F. J. Van Steenwijk, Am. J. Phys. **67** (1999) 486–492.

[29] M. Jeng, Am. J. Phys. **68**(1) (2000) 37–40.

[30] J. Cserti, Am. J. Phys. **68** (2000) 896–906.

[31] J. Cserti, G. David and A. Piroth, Am. J. Phys. **70** (2002) 153–159.

[32] Cserti J, Szechenyi G, and David G. 2011. J. Phys. A: Math. Theor. **44** (2011) 215201.

[33] J. H. Asad, A. Sakaji, R. S. Hijjawi and J. M. Khalifeh, Int. J. Theo. Phys. **43**(11) (2004) 2223–2235.

[34] J. H. Asad, A. Sakaji, R. S. Hijjawi and J. M. Khalifeh. (2006). *Eur. Phys. J. B*, (52) 2: 365.

[35] J. H. Asad, R. S. Hijjawi, A. J. Sakaji and J. M. Khalifeh, Int. J. Theo. Phys. **44**(4) (2005) 471–483.

[36] R. S. Hijjawi, J. H. Asad, A. J. Sakaji, M. Al-Sabayleh and J. M. Khalifeh, Eur. Phys. J. Appl. Phys. **41** (2008) 111–114.

[37] J. H. Asad, A. A. Diab, R. S. Hijjawi, J. M. Khalifeh. (2013). The European Physical Journal Plus. **128 (1)**: 1-5.

[38] Jihad H. Asad. Exact Evaluation of the Resistance in an Infinite Face- Centered Cubic Network. (2013). J. Stat. Phys. (accepted).

[39] G. Horwitz and H. B. CaUen, Phys. Rev. 124,1757 (1961).

[40] B. V. Thompson and D. A. Lavis, Proc. Phys. Soc. (London) 91,645 (1967).

[41] R. Tahir-Kheli and D. ter Haar, Phys. Rev. 127, 88 (1962).

[42] H. B. CaUen, Phys. Rev. 130, 890 (1963).

[43] N. W. Dalton and D. W. Wood, Proc. Phys. Soc. (London) 90, 459 (1967).

[44] E. W. MontroU, Nuovo Cimento Suppl. 6, 265 (1949).

[45] T. H. Berlin and M. Kac, Phys. Rev. 86, 821 (1952).

[46] M. Lax, Phys. Rev. 97, 629 (1955).

[47] W. F. Van Peijpe, Physica 5, 465 (1938).

[48] Watson, G. N. (1939). Three triple integral. Quarterly Journal of Mathematics (Oxford) 10, 266.